\documentclass[sigconf]{acmart}

\usepackage{epsfig}
\usepackage{subfigure}
\usepackage{calc}
\usepackage{multicol}
\usepackage{pslatex}
\usepackage{multirow}
\usepackage{booktabs}
\usepackage{tikz}
\usepackage{graphicx}
\usepackage{xcolor}
\usepackage{url}
\usepackage[T1]{fontenc}
\usepackage[utf8]{inputenc}
\usepackage{array}
\usepackage{pdfpages}

\usepackage[]{todonotes}

\newcolumntype{P}[1]{>{\centering\arraybackslash}p{#1}}
\newcolumntype{M}[1]{>{\centering\arraybackslash}m{#1}}

\AtBeginDocument{%
  }

\acmConference[ICSE 2024]{46th International Conference on Software Engineering}{April 2024}{Lisbon, Portugal}

\begin{document}

\title{An Actionable Framework for Understanding and Improving \newline Talent Retention as a Competitive Advantage in IT Organizations}

\author{Luiz Alexandre Costa}
\email{luiz.costa@edu.unirio.br}
\affiliation{%
  \institution{Universidade Federal do Estado do Rio de Janeiro}
  \city{Rio de Janeiro}
  \country{Brazil}
}
\author{Edson Dias}
\email{ecdias@ufpa.br}
\affiliation{%
  \institution{Universidade Federal do Pará}
  \city{Belém}
  \country{Brazil}
}
\author{Danilo Monteiro Ribeiro}
\email{danilo.ribeiro@zup.com.br}
\affiliation{%
  \institution{Zup Innovation}
  \city{São Paulo}
  \country{Brazil}
}
\author{Awdren Fontão}
\email{awdren@facom.ufms.br}
\affiliation{%
  \institution{Universidade Federal de Mato Grosso do Sul}
  \city{Campo Grande}
  \country{Brazil}
}
\author{Gustavo Pinto}
\email{gpinto@ufpa.br}
\affiliation{%
  \institution{Universidade Federal do Pará}
  \city{Belém}
  \country{Brazil}
}
\author{Rodrigo Pereira dos Santos}
\email{rps@uniriotec.br}
\affiliation{%
  \institution{Universidade Federal do Estado do Rio de Janeiro}
  \city{Rio de Janeiro}
  \country{Brazil}
}
\author{Alexander Serebrenik}
\email{a.serebrenik@tue.nl}
\affiliation{%
  \institution{Eindhoven University of Technology}
  \city{Eindhoven}
  \country{Netherlands}
}



\begin{abstract}
In the rapidly evolving global business landscape, the demand for software has intensified competition among organizations, leading to challenges in retaining highly qualified professionals in IT organizations.
One of the problems faced by IT organizations is the retention of these strategic professionals, also known as talent.
This work presents an actionable framework for talent retention in IT organizations called TR Framework. It is based on our findings from interviews conducted with 21 IT managers.
The TR Framework aims to guide IT managers in improving talent management processes by addressing specific challenges, identifying important factors, and exploring strategies at the individual, team, and organizational levels. The research findings shed light on the barriers faced by IT managers and the coping mechanisms employed when general improvement strategies prove ineffective. By fostering cohesive teams, promoting innovation, and ensuring knowledge continuity, talent retention becomes a significant competitive advantage for IT organizations. 
The TR Framework provides valuable insights and practical recommendations for enhancing talent retention and management practices to build successful software development teams.
\end{abstract}

\ccsdesc[500]{Human-centered computing~Collaborative and social computing}

\keywords{Talent Retention, Field Study, Work Motivation, Work Satisfaction}



\newcommand{\danilo}[1]{\textcolor{red}{ \textbf{Danilo:  }#1}}
\newcommand{\awdren}[1]{\textcolor{red}{ \textbf{Awdren:  }#1}}
\newcommand{\gnote}[1]{\textcolor{red}
{\textbf{[#1] - Gustavo}}}
\newcommand{\as}[1]{{\color{magenta}{Alexander: #1}}}
\newcommand{\luiz}[1]{\textcolor{blue}{ \textbf{Luiz:  }#1}}
\newcommand{\cmrdy}[1]{\textcolor{black}{#1}}
\newcommand{\rps}[1]{\todo[color=orange!40, inline]{\footnotesize{RPS: #1}}}
\maketitle

\section{Introduction} \label{sec:introducao}

In the contemporary global business scenario, the demand for software among companies to fulfill their objectives is increasing rapidly. 
This growing need for software has led to a significant increase in the number of software development companies, thus intensifying the competition for highly qualified IT professionals, as reported by Gartner Inc.\footnote{https://www.gartner.com/en/newsroom/press-releases/2023-04-06-gartner-forecasts-worldwide-it-spending-to-grow-5-percent-in-2023} However, this scenario is challenging due to the scarcity of individuals with specific technical skills, which represents a significant obstacle for organizations seeking to build highly proficient software development teams \cite{brynjolfsson2014second}.

One of the problems faced by IT organizations is the retention of their highly qualified IT professionals, also known as ``talent'' \cite{allen2010retaining, zhang2022turnover}. Fierce competition in the job market and a shortage of talent with specific technical skills lead to a high turnover rate in these companies \cite{allen2003role, devi2016impact}. The frequent loss of talent implies not only increasing operational costs but also the loss of accumulated knowledge and expertise, negatively affecting the efficiency of operations \cite{luftmann2008key}.

Moreover, it is important to point out that talent retention can become a significant competitive advantage for IT organizations, according to Resource-Based View (RBV) theory~\cite{barney1991firm}. Companies that manage to attract, develop, and retain highly qualified talent have the capacity to build cohesive and specialized teams capable of innovating, delivering high-quality results, and adapting quickly to technological changes \cite{bihani2014review}. In addition, talent retention promotes the continuity of knowledge and organizational expertise, providing stability and consistency in operations.

In this challenging context, talent management has become a critical factor for IT organizations \cite{phillips2008managing, hongal2020study}. Talent retention is not just limited to the initial attraction but also involves the adoption of efficient strategies and practices to keep highly qualified IT professionals engaged and satisfied in their work environment.
The job challenges and the continuous search for learning are specific factors for retaining talent in the area of Software Engineering (SE). These professionals are often driven by intellectual curiosity and a desire to tackle complex problems \cite{fraser2015reflections}. 
The impact for organizations of not retaining talent from an SE point of view may affect the company's competitiveness due to the loss of knowledge and expertise, recruitment costs, and the discontinuity of projects \cite{giuffrida2013empirical}.

To address this challenge and find solutions, we developed an actionable framework on talent retention in IT organizations called TR Framework.
It seeks to understand and guide IT managers in the improvement of the talent management process.
The TR Framework is grounded in data collected through interviews. We conducted interviews with 21 IT managers across different software industries, from small to large multinational corporations, to understand the challenges that their organizations face in retaining talent and emergent factors that are perceived as important to them. 
The interviews were performed from September 2022 to February 2023, during the COVID-19 pandemic and post-pandemic periods. In this period, we saw an increasing search for tech professionals in general \cite{koch2021looking, hylton2022long}.
Through our interviews, we also highlighted strategies IT managers use to overcome the barriers, as well as the coping mechanisms they turn to when general improvement strategies do not work.

The outcome of our work is a practical framework that accomplishes the following: i) presents a set of \textbf{factors} that may influence talent retention; ii) explores the \textbf{contextual characteristics} that may moderate the impact of the factors on talent retention; iii) identifies the \textbf{barriers} that cut across these factors and impede IT managers from improving their talent retention experience; and iv) documents the \textbf{strategies} and \textbf{coping mechanisms} employed by IT managers to overcome these barriers and enhance one or more dimensions of their talent retention experience.
Our framework emphasizes the idea that it is not only theoretical or conceptual but can also be effectively applied in practice to guide concrete actions and decision-making.

Studying IT managers is important for SE because they play a crucial role in overseeing software projects and managing teams of software engineers \cite{wiese2023managers}.
Through our results in this work, we hope to contribute to the advancement of SE by offering guidelines for talent management processes that can help IT organizations face the challenges of retaining talent in a highly competitive environment.


\section{Background \& Related Work} \label{sec:background}

\subsection{Defining and Contextualizing Talent}\label{subsec:definicao_talento}

The definition we use for \textbf{talent} in our work is: \textit{``talents are strategic people in the organization who have a unique combination of characteristics: i) they have the ability to exercise their occupation with technical excellence, mastering the necessary skills and knowledge; ii) they are aligned with the company's culture and business, sharing the organizational values and objectives; and iii) they are able to influence inside and outside their team, bringing positive impact and contributing to the success of the organization''}.

Our definition is inspired by Boudreau and Ramstad \cite{boudreau2005talentship} and Michaels et al. \cite{michaels2001war}. In the first one, the authors introduce the concept of ``talentship'' and discuss the importance of talent segmentation in organizations. They emphasize the need for organizations to identify and nurture individuals with unique combinations of skills and characteristics to achieve sustainable success. In the latter, the authors discuss the fierce competition for highly skilled individuals in various industries, including IT. 
The authors highlight the characteristics that define talented people.

We also use the concept of \textbf{talent retention}.
The definition of this concept used in our work is \textit{``an organizational process that aims to keep the best employees in the organization, called `talent'}''. Therefore, talent retention focuses on the retention of strategic individuals within an organization.
It is important to note that other similar concepts, such as employer retention and turnover intention, differ from what we refer to as talent retention.

Employer retention refers to the ability of an organization to encourage employees to remain in the organization for a long period~\cite{schyns2007turnover}, without necessarily focusing on talent-specific factors. It encompasses strategies and practices aimed at reducing overall employee turnover within the organization.
On the other hand, turnover intention relates to an individual employee's intention or inclination to change the job or organization voluntarily~\cite{das2013employee}. 

\subsection{Talent Retention} \label{subsec:talento_vantagem_competitiva}

The studies in the talent retention field often relate talent retention to other human factors such as motivation, job satisfaction, and leadership. For instance, Ghani et al. \cite{ghani2022challenges} conducted a literature review on challenges and strategies for talent retention in the healthcare sector. The authors organized their findings based on equity 
theory \cite{adams1976equity}, job characteristics theory \cite{hackman2015job}, expectancy theory \cite{vroom2015expectancy}, and Maslow's theory \cite{vroom2015expectancy}. 
The authors link factors from these theories, such as career growth, leadership, communication, trust in the organization, and work relationships, to employee satisfaction, which ultimately impacts talent retention. The study also presents strategies for talent retention in the healthcare sector, such as implementing work-hour policies, offering recognition, bonuses, and other retention strategies.

Effective talent management is critical for IT organizations, seeking to gain a competitive advantage in a highly competitive market \cite{michaels2001war, hadijah2023implementation}. This perspective highlights the strategic importance of identifying, recruiting, and retaining top IT professionals amidst the competitive talent scenario.
It offers valuable insights into how organizations can gain a competitive edge through efficient talent management. The authors address the unique challenges IT organizations encounter in talent retention and provide best practices for fostering an environment that promotes the growth and involvement of talented professionals.

This approach is aligned with RBV theory \cite{barney1991firm}. The RBV is a widely recognized theory in the field of business strategy and management \cite{barney2001resource, ray2004capabilities}. It focuses on analyzing an organization's internal resources as a source of competitive advantage.
According to the RBV, an organization's unique resources and capabilities, such as tangible  (e.g., infrastructure, technology) and intangible assets (e.g., knowledge, organizational culture) can be sources of sustainable competitive advantage. The theory emphasizes that the combination and effective use of these resources can generate value for the organization and differentiate it from competitors.


When considering the talent retention landscape \cite{vaiman2011talent, hongal2020study, allen2021global}, it is crucial to take specific factors (i.e., economic uncertainties, social inequalities, limited supply of technical specialists, and cultural and social aspects) into account. Understanding local dynamics and challenges helps IT organizations adopt more effective strategies to retain talent and gain sustainable competitive advantage.

\subsection{Talent Retention in Software Engineering}

Talent retention management is a widely studied topic in literature, and it applies to various industries and employee categories \cite{cappelli2008talent}. However, software engineers have unique characteristics and challenges that can differentiate them from other employees \cite{turley1995competencies, uden2004lifelong, francca2018motivation}. To cite a few, innovation and knowledge continuity, in which software engineers thrive on innovation and continuous learning, and the changing technology scenario that requires software engineers to stay updated with the latest tools.

Finally, the COVID-19 pandemic has presented significant challenges for IT software producing organizations regarding talent retention \cite{aguinis2021talent}. With the rapid transition to remote work and the economic uncertainty resulting from the crisis, many IT professionals have found themselves reassessing their career priorities, which has led to an increase in turnover \cite{shen2020impact}. The balance between personal and professional lives has become more complex \cite{russo2021daily}. IT organizations face the challenge of adopting flexible retention strategies to meet the needs and expectations of employees and to keep their talent engaged and motivated during this period \cite{bailey2021covid}.


\section{Research Method} \label{sec:metodologia}

We ground our work on Empirical Software Engineering (ESE) guidelines~\cite{stol2020guidelines}. 
We conducted a field study as a research method to identify specific characteristics of the talent retention process in IT organizations, such as culture, hierarchical structure, workload, human resource policies, and benefits. Furthermore, it provides an opportunity for participants to share their experiences and perspectives based on semi-structured interviews, as proposed by Seaman \cite{seaman1999qualitative}. 
To do so, we performed a set of interviews based on the recommendations for field studies \cite{singer2008software}.
Although the subject of talent retention has been widely studied in management literature \cite{cappelli2008talent}, we sought to obtain clarification through interviews on what makes software engineers different from any other employee.
The participants were composed of a diverse set of IT managers in terms of role, industry, and experience \cite{myers2002qualitative}.

Next, we used qualitative analysis to develop a framework to understand the talent retention process and identify key drivers and interactions between relationships emerging from the collected data. We applied coding procedures to analyze qualitative data, inspired by the initial procedures for the grounded theory of Strauss and Corbin \cite{strauss1990basics}.
In this section, we provided an overview of our interview participant selection process, the questions used, and our approach for analyzing the results.

\subsection{Research Questions} \label{subsec:rq}


In line with our research method, our research questions aimed to allow the researcher to obtain detailed and rich information about participants' experiences, opinions, and perspectives on the topic at hand.  The research questions evolved as we collected and examined our data, and they were further revised.

Our initial research question was: \textbf{What physical, human, and organizational factors drive talents to \emph{stay} in IT organizations?}
As we conducted our interviews, we not only uncovered factors that matter in talent retention to IT managers but also found contextual characteristics of their work that promote talent retention in the organization.
Respondents also shared barriers impeding their ability to retain talents, strategies for successfully keeping talents on their teams, and challenges in implementing innovative mechanisms to attract talents. 
We also identified coping mechanisms to overcome difficulties in retaining talented software engineers by IT managers. 
Our work begins with ``initial research question'', which evolve throughout the work \cite{charmaz2006constructing, greiler2022actionable}. Thus, the emergent research questions (RQ) and its detailed as follows:

\noindent
\textbf{RQ1:} Which important \textbf{factors} influence talent retention in IT organizations? Here our goal is to understand the factors that influence a talent's decision to stay (or leave), as well as obtaining valuable insight into employees' concerns and motivations. 

\noindent
\textbf{RQ2:} Which \textbf{contextual characteristics} moderate how important a factor is talent retention’s experience in the organizations? Here our goal is to identify external factors that may moderate organizations' ability to retain talent;

\noindent
\textbf{RQ3:} Which \textbf{barriers} prevent IT managers from improving factors that affect talent retention? By identifying the barriers, our goal is to improve the factors that affect the permanence of a talent, such as financial resources, bureaucracy, and skills.

\noindent
\textbf{RQ4}: Which \textbf{strategies} do IT managers employ to improve talent retention in their organizations? Knowing the practices that IT managers are using to retain talent could potentially reduce employee turnover.

\noindent
\textbf{RQ5:} Which \textbf{coping mechanisms} do IT managers resort to when factors that negatively impact talent retention are not improved? Finally, our goal here is to identify gaps in talent retention policies beyond existing efforts.


\subsection{Semi-Structured Interviews} \label{subsec:semi-structured_interviews}

Through semi-structured interviews with 21 IT managers, we set out to gather opinions on i) the factors that affect the permanence of talent in the organization \textit{(RQ1)}; ii) the organizational environment characteristics that can modify the importance of a factor in the talent retention experience \textit{(RQ2)}; iii) the barriers and challenges that managers face in retaining talent \textit{(RQ3)}; iv) the strategies that managers use to retain talent \textit{(RQ4)}; and v) the turnaround mechanisms that managers use to ensure talent retention actions \textit{(RQ5)}.

The interviews were conducted in Portuguese and were performed from September 2022 to February 2023, during the COVID-19 pandemic and post-pandemic periods. The interviews lasted an average of 38 minutes. We used Zoom to conduct and record each interview. We also transcribe each recording with transcription software. 
We followed the interview questions from the interview guide which can be found at \textit{[link omitted]}.
As a part of the field study method, the semi-structured interviews helped bring theory closer to practice and provided a deeper understanding of the real-world challenges that IT managers face \cite{seaman1999qualitative}.

In the first part of the interviews, we provided each participant with a high-level definition of talent, as described in Section \ref{subsec:definicao_talento}.
In the second part, we asked about each participant's professional history, age, and academic background. The intention was to bring the participant closer to the interviewer, making him more comfortable during the interview. 
Finally, in the third part of the interviews, we focused on understanding how participants could improve talent retention in their teams, the importance of investing in talent retention, the IT organizations' strategies and policies for retaining their most talented and valuable employees, and the difficulties or barriers that managers may run across when attempting to improve the elements that influence a talent's durability.


\subsection{Characterization of Participants} \label{subsec:interview_participants}

To select participants, we used convenience sampling by reaching out to IT managers in our network. We used email and other online communication channels (WhatsApp and LinkedIn) to reach them.
Our selection criteria to recruit credible practitioners was based on guidelines of Rainer and Wohlin \cite{rainer2022recruiting}:
i) relevant experience: IT managers have knowledge and practical experience, including leadership skills, recruiting, and retaining talent; ii) strategic position: IT managers are directly involved in making decisions related to personnel management and play an important role in retaining talent within their teams; iii) variety of organizations: by selecting IT managers from different types of IT organizations makes it possible to get a broader picture of the situation; and iv) diversity of perspectives: by selecting IT managers from different backgrounds, genders, experience levels, and professional backgrounds, we can get different points of view on the topic of talent retention. 

Prior to each meeting, we asked participants for their consent to be interviewed and for their permission to record the session. We informed participants that they could withdraw from the interview at any time and that their responses would be deleted.
All participants have management skills and experience in matters related to talent retention, which helps ensure that the selected sample is representative and relevant to the research objectives.
Six of the 21 participants had less than ten years of IT management experience, and nine of those had 15+ years of experience. Four of the participants had between five and seven years of experience. Participants worked in a variety of industries (including the insurance sector, auditing, government banking, fintech, and general software consulting). The size of the teams our participants manage ranged from 6 to 250 people, and their company sizes varied between 80 and 49K+ people.
Table \ref{tab:participantes_entrevistas} shows a summary of the IT managers we interviewed. 


\begin{table}[ht!]
\footnotesize
\centering
\caption{Details of participants from 11 different companies.}
\label{tab:participantes_entrevistas}
\begin{tabular}{p{0.01\textwidth} p{0.04\textwidth} p{0.07\textwidth} p{0.02\textwidth} p{0.10\textwidth} p{0.06\textwidth}}
\toprule
ID & Company Size & Industry & Team Size & Current Role & Experience
\\ \midrule
P1 & 100 & Software Consulting & 15 & IT Manager & 6 yrs
\\ \midrule
P2 & 4,000 & Software Consulting & 105 & Head of IT \newline Operations & 29 yrs
\\ \midrule
P3 & 7,000  & Insurance & 35 & IT Manager & 26 yrs
\\ \midrule
P4 & 10,000  & Auditing & 32 & IT Manager & 5 yrs
\\ \midrule
P5 & 49,000  &  Government Banking & 12 & IT Manager & 10 yrs
\\ \midrule
P6 & 7,000 & Insurance  & 6 & Tech Lead & 15 yrs
\\ \midrule
P7 & 400  & Business \newline Intelligence & 9 & Data \newline Engineer Lead & 6 yrs
\\ \midrule
P8 & 3,000  & Software Consulting & 70 & IT Manager & 5 yrs
\\ \midrule
P9 & 3,000  & Software Consulting & 92 & IT Director & 12 yrs
\\ \midrule
P10 & 3,000  & Software Consulting & 60 & IT Manager & 8 yrs
\\ \midrule
P11 & 200  & Software Consulting  & 80 & Chief Information Officer & 34 yrs
\\ \midrule
P12 & 4,000  & Software Consulting  & 250 & Vice President & 30 yrs
\\ \midrule
P13 & 7,000  & Insurance  & 15 & IT Manager & 20 yrs
\\ \midrule
P14 & 1,200  & Software Consulting  & 38 & IT Manager & 7 yrs
\\ \midrule
P15 & 15,000 & Software Consulting  & 30 & IT Manager & 30 yrs
\\ \midrule
P16 & 7,000  & Insurance  & 8 & IT Manager & 32 yrs
\\ \midrule
P17 & 4,000  & Software Consulting & 150 & Director of \newline IT Operations & 15 yrs
\\ \midrule
P18 & 7,000  & Insurance  & 19 & IT Manager & 23 yrs
\\ \midrule
P19 & 3,000 & Software Consulting  & 35 & IT Manager & 10 yrs
\\ \midrule
P20 & 3,000 & Software Consulting  & 15 & IT Manager & 10 yrs
\\ \midrule
P21 & 80  & Fintech  & 80 & Founder & 13 yrs
\\ \bottomrule
\end{tabular}
\end{table}

\subsection{Interview Procedures} \label{subsec:interview_process}

Before refining our interview guide and conducting the final set of 21 interviews, we performed a pilot with three interviewees (managers). We did not use the data from the pilot research in our analysis.
The pilot encouraged us to add a definition of talent retention at the beginning of the interview.  Without a clear definition of talent, there may be different interpretations of who is considered a talent in the  organization, which could lead to divergent responses. It also helped us ensure that everyone was consistently understanding the questions and gave us the confidence to improve the interviewing process. 

The concept of ``saturation'' was adopted to establish the number of required interviews. According to Creswell \cite{creswell2016qualitative}, a saturation is reached when performing a new set of interviews does not come up with any new emerging data. It is a challenge to define how many interviews are required to conclude a field study. We exceeded the reference recommendation of Guest and colleagues \cite{guest2006many}, who explain that a saturation usually could be obtained with 12 interviews.

Due to a lack of new data as we go through the interviews, we observe whether participants are repeating information and topics already discussed earlier. Thus, we realized that our codes and insights were fully saturated \cite{glaser1999theoretical}. Interview recordings and transcriptions were continually revisited in an iterative process. For each new interview, we went back to previous interviews to see if previous interviewees also mentioned the new insights. Once no new codes or insights emerged in three consecutive interviews, we concluded that our findings were saturated and stopped recruiting new participants.

\subsection{Coding Process} \label{subsec:coding_process}

To analyze the interviews, we performed an open coding approach where we coded the interviews inspired by the initial procedure for the grounded theory of Strauss and Corbin \cite{strauss1990basics}. The method uses an inductive approach (bottom-up), which means that theories emerge from collected data rather than being imposed by the researcher \cite{charmaz2006constructing}.

Two researchers conducted and coded the interviews over an average of three iterative cycles. The codes were generated inductively, i.e., emerged directly from the data without prejudice or prior assumptions by the researchers. To ensure a collaborative, comprehensive, and reliable approach, all interviews were reviewed three times by each researcher to allow for the exchange of perspectives and consensus on code generation. During the coding process, the points of disagreement between researchers were discussed transparently and resolved through consensus in a new review meeting. Three other researchers with more than 15 years in ESE double-checked the results and ensured the compliance of the final dataset.

The two researchers divided the \textbf{transcripts} of the participants into coherent units (sentences or paragraphs) and added \textbf{preliminary codes} that represented the key points that each participant talked about. Based on the preliminary codes, we set the \textbf{focused codes} that captured the most frequent and relevant factors affecting talent retention, as shown in Table \ref{tab:code_example}.
Next, we formed axial coding as described by Charmaz \cite{charmaz2006constructing} to group the codes into categories using Atlas.TI tool. 
We also perform three iterative cycles with discussions among the researchers to write memos for the codes and categories and build relationships across the codes, forming a logical network to a higher degree, as shown in Table \ref{tab:code_example}.

\begin{figure}[t!]
    \centering
    \includegraphics[scale=0.6, clip=true, trim=60px 290px 30px 200px]{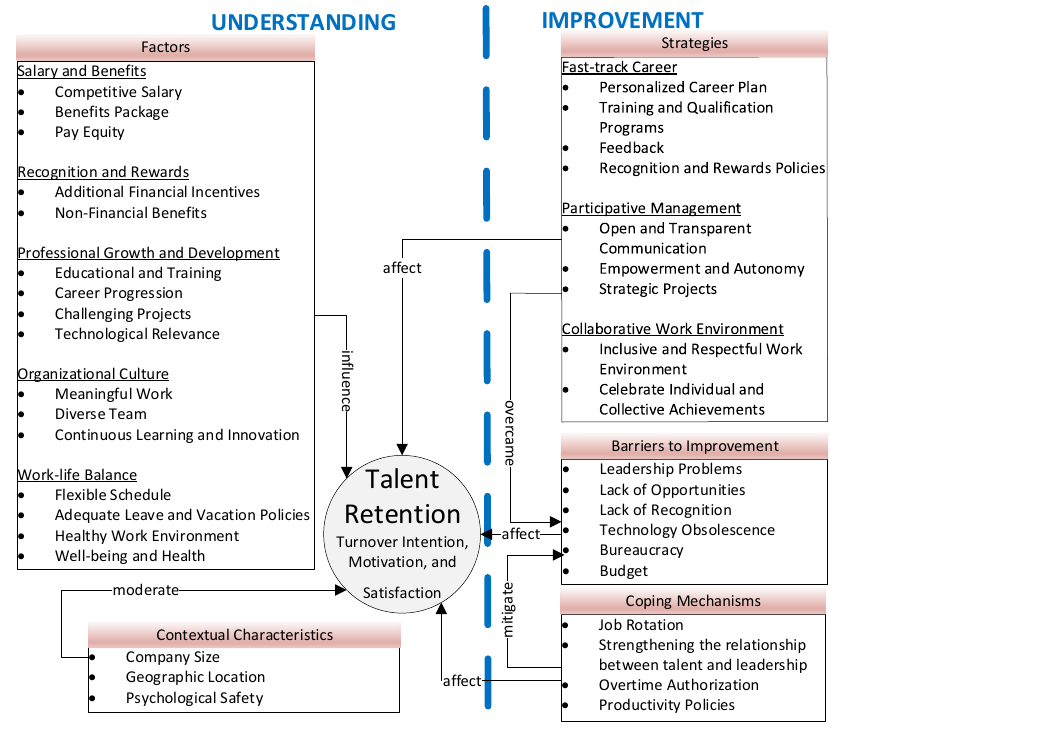}
    \caption{TR Framework.}
    \label{fig:framework}
\end{figure}

In our analysis, the TR Framework was built, consisting of five core categories (that may aggregate other categories in our code hierarchy), inspired by the work of Greiler \textit{et al.} \cite{greiler2022actionable}: 1) \textbf{factors} that influence the retention of talent in organizations; 2) \textbf{contextual characteristics} that encompass the particularities of the environment and moderate the importance of factors; 3) \textbf{barriers} that identify the obstacles that prevent IT managers from improving talent retention; 4) \textbf{strategies} that address the actions and initiatives of IT managers to improve talent retention; and 5) \textbf{coping mechanisms} that IT managers resort when they face difficulties in improving the talent retention experience in the organizations.
These five core categories are key components in the TR Framework that emerged from our analysis, as shown in Figure \ref{fig:framework}. The TR Framework is our main research outcome and it also helped us refine our preliminary research questions as presented in Section \ref{subsec:rq}.
The core categories and codes will be described in Sections \ref{sec:resul_understanding} and \ref{sec:resul_improving}.
Table \ref{tab:code_example} shows examples of the coding process for some transcripts, the resulting codes, and core categories.

\begin{table*}[h!]
\centering
\caption{Illustration of the coding process.}
\label{tab:code_example}
\begin{tabular}{@{}|llll|@{}}
\toprule
\toprule
\multicolumn{4}{|l|}{\begin{tabular}[c]{@{}l@{}}\textbf{Transcript Unit:} Money is a decisive factor in your decision to look for a new opportunity, but it is not the main one. It depends on \\the context in which you work. For example, I started as a programmer, progressed in all steps of my career in the technical area, and \\then moved on to the management area. I remember that what most motivated me was constant learning and technological innovation.\\ I also see this happening nowadays with talented people. One of the things that most motivates talented people is constant learning.\\ While they are learning and gaining knowledge about a new technology that will increase their employability in the future, the tendency \\is for them to stay with the company. \textbf{(P12)}\end{tabular}} \\ \midrule
\multicolumn{1}{|l|}{\begin{tabular}[c]{@{}l@{}}\textbf{Preliminary Code:}\\ Talented people are motivated when they work\\ in a constantly learning and innovative environment.\end{tabular}} &
\multicolumn{1}{l|}{\begin{tabular}[c]{@{}l@{}}\textbf{Focused Code:}\\Continuous Learning and\\ Innovation\end{tabular}} &
\multicolumn{1}{l|}{\begin{tabular}[c]{@{}l@{}}\textbf{Category:}\\ Organizational\\ Culture\end{tabular}} &
\multicolumn{1}{l|}{\begin{tabular}[c]{@{}l@{}}\textbf{Core Category:}\\ Factors\end{tabular}} \\ 

\toprule
\toprule
\multicolumn{4}{|l|}{\begin{tabular}[c]{@{}l@{}}\textbf{Transcript Unit:} We have a slow process of effecting salary increases, so it sometimes causes us to lose talent. I have to wait for internal \\and international bureaucracies. \textbf{(P17)}\end{tabular}} \\ \midrule
\multicolumn{1}{|l|}{\begin{tabular}[c]{@{}l@{}}\textbf{Preliminary Code:}\\Bureaucracy is a barrier to improving talent retention.\end{tabular}} &
\multicolumn{1}{l|}{\begin{tabular}[c]{@{}l@{}}\textbf{Focused Code:}\\Bureaucracy\end{tabular}} &
\multicolumn{1}{l|}{\begin{tabular}[c]{@{}l@{}}\textbf{Category:}\\-\end{tabular}} &
\multicolumn{1}{l|}{\begin{tabular}[c]{@{}l@{}}\textbf{Core Category:}\\ Barriers to Improvement\end{tabular}} \\

\bottomrule
\bottomrule
\end{tabular}
\end{table*}

\section{TR Framework} \label{sec:framework}

The primary result derived from our work is the \textbf{TR Framework}, as shown in Figure \ref{fig:framework}. The core concept within our framework is \textbf{T}alent \textbf{R}etention, characterized by satisfaction and motivation as mind dimensions. The other two parts of the framework are called \textit{understanding} (left side) and \textit{improvement} (right side).

The framework's division into ``understanding'' and ``improvement'' helps organize proposed actions into a logical flow, where in-depth understanding serves as the foundation for formulating relevant, actionable improvements. On the left side, we have the two categories of factors that IT managers shared that impacted their experience and contextual characteristics that moderated the importance of these factors to the IT managers. On the right side, we have three other essential categories focusing on enhancing talent retention. These categories include barriers, strategies, and coping mechanisms that IT managers employ when strategies do not work. We describe all categories in Sections \ref{sec:resul_understanding} and \ref{sec:resul_improving}.
\section{Understanding Talent Retention} \label{sec:resul_understanding}

\subsection{Factors Affecting Talent Retention}
\label{subsec:factors}

We aimed to identify key elements that IT managers perceive as the most important factors affecting talent retention in IT organizations. As described in Section \ref{sec:metodologia}, we iteratively coded their responses to RQ1 and grouped them into categories. These categories represent themes that helped us understand a set of factors as a group. Identifying and understanding these factors will help us create  practical retention strategies.

We discuss in the following section the emerged TR factors, our focused codes for this core category, and the categories we grouped them through axial coding.
We did not count how many times each factor emerged, as the interviews were open-ended, and any counting could be misleading. However, two or more participants mentioned all the factors that emerged. For the sake of research transparency, we have preserved a mapping of participants to factors in our raw data, which is available online \textit{[link omitted]}.

\subsubsection{Salary and Benefits}
\label{subsubsec:remuneracao}

The focused codes in this category were the most cited by the participants regarding talent retention in IT organizations. 
\textbf{Competitive salary} is essential to attracting and retaining qualified talent. IT professionals usually have specialized skills and are in high demand in the market. Therefore, offering salaries that are competitive with the industry average is important to ensure that employees are not tempted to seek better opportunities elsewhere. In addition to the base salary, the \textbf{benefits package} is also significant. It may include benefits such as healthcare assistance, life insurance, a pension plan, or food vouchers. Offering attractive benefits can increase employee satisfaction and demonstrate the organization's commitment to their well-being and quality of life. 
As P4 shared: \textit{``The main factor is remuneration, money really, even more so when you talk about IT, where we see very aggressive startup companies with very competitive salary and benefits''.}

\textbf{Pay equity} is another important factor in talent retention. IT employees expect their compensation to be fair in relation to their responsibilities, experience, and performance, as well as compared to their peers. It is essential to ensure a fair and transparent remuneration policy, avoiding unjustified wage disparities that could lead to dissatisfaction and the loss of talent. 
As P3 commented: \textit{``I would not want to leave here, saying: look, if you give flowers every morning, people will not leave. Just imagine that when IT professionals realize they are being paid fairly and equally with colleagues in the same role, it creates harmony in the work environment''}.

\subsubsection{Recognition and Rewards} \label{subsubsec:recompensa}

The focused codes included in this category refer to the appreciation of employees' work. It is critical to the motivation and engagement of IT professionals.
\textbf{Additional financial incentives} can also be an effective form of retention. It may include performance bonuses, profit sharing, stock option programs, or other results-based incentives. These types of incentives can motivate employees to reach goals and reward good performance. 
P2 stated: \textit{``Many IT professionals come to me precisely to complain about the lack of recognition, not from the immediate management but from the company as a whole. In most of these cases, we managed to get some kind of reward to do the retention, but it was reactive''}.


In addition to financial compensation, there are \textbf{non-financial benefits} that can be equally valuable for retaining IT talent. This can include learning and development opportunities, mentoring programs, flexible schedules, a pleasant work environment, an inclusive organizational culture, recognition, and non-monetary rewards such as additional days off, wellness programs, remote work options, prizes, accolades, and opportunities for growth. 
As P9 shared: \textit{``I try to provide outstanding experiences, such as children's birthdays. You create an emotional bond, right? If one day these talents leave the company, they will be missed''}.

\subsubsection{Professional Growth and Development} \label{subsubsec:desenv_profissional}

It refers to the possibilities offered to IT professionals to advance their careers, acquire new skills, and expand their knowledge.
\textbf{Education and training} opportunities are essential to the professional development of IT employees. It may include participation in specific courses, workshops, seminars, webinars, and training programs to update technical skills. These initiatives allow IT professionals to stay current with the latest technologies and work practices, improving their effectiveness and value in the marketplace.
P13 stated: \textit{``Investment is associated with talent retention. Not only in terms of salary increases but also in terms of investment in training to place talented people in positions where they can make the difference they have''}.


\textbf{Career progression} 
may involve defining a career plan with clear steps and the possibility of promotions based on performance and professional growth. Promoting continued employee development is achieved by offering challenging goals and the opportunity to take on additional responsibilities.
P7 stated: \textit{``Career progression is kind of the fuel that feeds IT professionals' desire for growth. When they see clear opportunities for development in the company's hierarchy, they are more engaged to stay in the same place''}.

Enabling IT professionals to take on \textbf{challenging projects} is an effective way to promote their growth and development. By involving them in innovative and complex projects, the practitioners have the opportunity to expand their skills, face new challenges, and learn from practical experiences. 
As P13 shared: \textit{``Money is important, but having a challenging work proposal is essential to retaining talent''}.
Having \textbf{technological relevance} is critical to IT professional development. It may include access to technical libraries, software development tools, test labs, online learning platforms, and other sources of knowledge and support. Technological relevance indicates that a product or technology is current, useful, and in line with current standards and demands. IT professionals generally seek to work with modern and innovative technologies. A company that stays up-to-date technologically increases the satisfaction of IT professionals.
As P5 shared: \textit{``The other day, an employee told me: when we work on a software project with new technology, it becomes motivating to stay in the company''}.

\subsubsection{Organizational Culture} \label{subsubsec:cultura_organizacional}

A solid and positive organizational culture can create a sense of identification and connection between IT professionals and the organization, becoming a way to retain talent. The factors included in this category refer to values, beliefs, norms, practices, and behaviors shared within an organization. IT professionals value the opportunity to do \textbf{meaningful work} where they can rewardingly apply their skills and knowledge.
As P18 stated: \textit{``The reason for a talent's decision not to stay in the company was the work proposal and the daily routine that were not compatible with what he wanted. He no longer had a sparkle in his eyes''}.


A \textbf{diverse team} in an organizational culture that values diversity and inclusion can be attractive to IT professionals as it fosters innovation, creativity, and mutual learning. 
P4 shared: \textit{``Our company is proud to be a diverse and inclusive workplace, with people from several ethnicities and genders''}.
A culture that values \textbf{continuous learning and innovation} is attractive to IT professionals. Talent seeks organizations that encourage professional growth, offer development opportunities, foster creativity, and value experimentation and continuous improvement. A culture that encourages learning and innovation encourages talent engagement and retention. 
P12 stated: \textit{``One of the things that most motivate talented people is constant learning. While they are learning and gaining knowledge in a new technology that will increase their employability in the future, the tendency is for them to stay with the company''}.

\subsubsection{Work-life Balance} \label{subsubsec:equilibrio_vidapessoal_profissional}

\textit{Work-life balance} refers to IT professionals' ability to balance their work responsibilities and demands with their personal needs and commitments outside of the work environment. The factors addressed in this category play a crucial role in talent retention as they directly influence employee well-being, satisfaction, and quality of life. \textbf{Flexible schedules} are an effective way to promote work-life balance. This could include flexible work schedules, compressed schedules, part-time work schedules, or the option to work remotely. This flexibility allows IT professionals to accommodate their personal responsibilities, such as taking care of the family, carrying out leisure activities, or dealing with important personal commitments.
As P14 shared: \textit{``We place great importance on fulfilling our activities with quality and within the established deadlines. Other than that, we are flexible with regard to hours''}.

A \textbf{healthy work environment} is essential to ensuring a balance between personal and professional lives. It may involve implementing stress-reducing practices such as encouraging regular breaks, promoting a balanced work culture, and stimulating mutual support among team members. Additionally, encouraging off-hours digital disconnection and avoiding excessive workload are important steps to ensure employees have the time to take care of their personal needs.
P10 stated: \textit{``We lost a talent in our team due to a hostile environment with a lot of pressure for fast and quality deliveries''}.

Encouraging employee \textbf{well-being and health} is another factor in promoting work-life balance. This can include wellness programs such as i) physical activity, meditation, or yoga sessions; ii) access to counseling services or psychological support; and iii) incentives to adopt healthy habits. IT professionals who invest in their health and well-being contribute to their quality of life and overall happiness, making them more likely to stay with the organization.
As P20 shared: \textit{``We offered him a good salary increase, but even so, he wanted to leave the company. He wanted to work fewer hours to have a better quality of life''}.

\subsection{Contextual Characteristics} \label{subsec:contextual}

When considering the core category \textit{contextual characteristics}, it is important to adapt retention strategies to meet the specific needs of the IT organization and the environment in which it operates. Every organization has its own unique context, and understanding these elements is critical to implementing effective approaches to overcoming barriers during the talent retention process.
These contextual characteristics are outlined below, where we show them as moderating the impact of factors on the talent retention experience by IT managers.

\textbf{Company size} can play an important role in retaining talent. Companies of different sizes can offer different advantages and challenges for IT professionals. For example, in smaller companies, IT professionals may have the opportunity to assume a variety of responsibilities and enjoy a closer and more collaborative atmosphere. On the other hand, larger companies can offer more robust resources and opportunities for growth. Understanding company size and its implications can help tailor retention strategies to specific needs.
As P6 described: \textit{``I lost a great talent to a startup -- a small, versatile, and innovative company that promised to become a unicorn\footnote{Privately held startup company with a valuation of US\$1 billion or more.}. This employee chose to leave based on this scenario and still gained the advantage of having his position evolve much higher''}.


The \textbf{geographic location} of the organization also plays a significant role in retaining talent. In some regions, there may be a greater concentration of IT job opportunities, while in others, there may be talent shortages. In addition, factors such as quality of life, access to services, cost of living, and technological infrastructure can influence the attractiveness of a region for IT professionals. 
As P14 explained: \textit{``In a country such as [omitted], some talents have the will to work in a more promising IT market or in interior regions. Others want an international career for salary appreciation in another currency. I had a talent who was assigned to an international project so that he could reside in the country he desired. Currently, in the post-pandemic scenario and with the possibility of home office, it has not happened''}. 



Lastly, participants mentioned \textbf{psychological safety}, which refers to an environment in which people feel safe to express their opinions, ideas, and concerns without fear of reprisal or negative judgment. According to Edmondson \cite{edmondson1999psychological}, there is a strong relationship between psychological safety and team performance, noting that a climate of safety promotes the open exchange of information, knowledge sharing, and problem solving more effectively. P20 reinforced: \textit{``I started to work with the team to understand each other's career goals. I think this approach made people more motivated and engaged. This closer and more frequent career follow-up provided a more trusting work environment''}.

\section{Improving Talent Retention} \label{sec:resul_improving}

\subsection{Barriers to Improving Talent Retention} \label{subsec:barriers}

Barriers are obstacles or challenges that prevent the retention of talent in IT organizations. Once IT managers identify the barriers, the management team will be aware, actively work to overcome them, and develop mechanisms in order to improve retention.

Leadership plays a key role in retaining talent. IT managers who demonstrate a lack of leadership skills, a lack of support for employees, or toxic behaviors can undermine talent satisfaction and engagement. The presence of \textbf{leadership problems} can discourage talent from staying with the organization and create an unfavorable work environment. 
As P15 shared:  \textit{``Empathy with the immediate manager is fundamental for the continuity of the work. Before taking an administrative decision, I try to talk to the person to find out what is going on, mainly related to the drop in performance''}. 

The \textbf{lack of opportunities} for professional advancement makes IT professionals look to the market for opportunities for growth and development in their careers. If the organization does not offer clear opportunities for professional advancement, such as training, qualification courses, challenging projects, or promotions, talents may feel stagnant and decide to look for other opportunities in organizations that offer these growth perspectives.
P1 shared a personal experience:  \textit{``In an internal interview, the department manager was very clear and told me: look, here in this corporate area, you can only get to this point. So I declined the invitation to change and started to look for an opportunity in the market''}.

Recognition is an important motivational factor for employees. If IT managers do not recognize talent's contributions and accomplishments, it can lead to a feeling of worthlessness. The \textbf{lack of recognition} can lead talent to seek recognition and rewards in other organizations.
P1 stated: \textit{``Due to the fact that the IT area is very dynamic, personal issues are not noted as much. The IT area is much more focused on technical bias. The lack of recognition is a factor that encourages talent not to stay in the organization''}.


The rapid evolution of technology is a reality for IT organizations. If the organization fails to keep up with or provide talented people with access to modern and relevant technologies, it can lead to \textbf{technology obsolescence}. Lack of up-to-date resources and outdated technologies can make work less challenging and interesting for IT professionals, reducing their motivation and satisfaction. In addition, talent can look for organizations that offer opportunities to work with more advanced and innovative technologies. 
As P5 shared: \textit{``Legacy system does not motivate anyone. You only take care of problems. When we work on legacy systems with obsolete technologies, it does not encourage anyone to stay''}.

The \textbf{bureaucracy} can create obstacles and delays in decision-making processes, implementing changes, and accessing necessary resources. This can be especially problematic in IT organizations, where agility and quick responsiveness are highly valued. Excessive bureaucracy can discourage talent, who often seeks an agile and flexible environment, from remaining in the organization. Furthermore, bureaucracy can create barriers to innovation and professional advancement, limiting growth opportunities for talented people.
As P17 stated: \textit{``We have a slow process of effecting salary increases, so it sometimes causes us to lose talent''}.

When \textbf{budget} is constrained, it can be difficult for the organization to compete with other companies that offer more attractive compensation packages. Furthermore, a lack of financial resources may limit the organization's ability to provide adequate incentives and rewards.
As P7 shared: \textit{``We have a budget to carry out salary adjustments. However, it is possible to have an exception that exceeds this limit. In this case, we must escalate to the IT board.''}

\subsection{Improvement Strategies} \label{subsec:strategies}

This core category involves formulating concrete strategies to address talent retention. Based on our findings
we identify specific strategies to attract, engage, and retain talent in IT organizations. 

\subsubsection{Fast-track Career} \label{subsubsec:fasttrack_career}

Items covered in this category indicate a desire to progress quickly in an IT career, achieve promotions, or move faster in terms of job responsibilities and opportunities. 
Creating a \textbf{personalized career plan} that provides opportunities for growth, such as challenging projects, job rotation, participation in strategic teams, project leadership, and international career opportunities.
P17 stated: \textit{``To be able to retain talent, it is necessary to understand what each person has as a professional desire. The HR technical recruiter interviews all developers every 3 months. Then we generate reports and graphs to identify what they worked with and what they would like to work with. We establish a set of priorities together. So, it is possible to propose an individualized career plan''}.

Investment in the development of talents through \textbf{training and qualification programs}, such as internal or external courses, workshops, mentoring, and continuous learning opportunities, can help to improve the skills and knowledge of talents, promoting their professional growth and increasing their job satisfaction. 
As P3 shared: \textit{``We have a partnership on an online course platform for employees''}.
Provide constructive \textbf{feedback} on a regular basis, identify areas for improvement, and offer guidance so talented people can grow and progress in their careers.
As P1 explained: \textit{``Give constant feedback to align objectives and avoid understanding noise''}.

\textbf{Recognition and rewards policies} aim to recognize and reward exceptional performance. It may include formal or informal recognition programs such as public accolades, awards, or performance-based bonuses. In addition, offering opportunities for career growth and promotions can also be a way to reward talent and encourage them to stay with the organization. 
As P18 shared: \textit{``A form of recognition that does not require financial investment but makes people feel valued is to highlight or praise in corporate groups''}.

\subsubsection{Participative Management} \label{subsubsec:participative_management}

Strategies in this category describe a leadership style in which employees are encouraged to actively participate in organizational decisions, promoting collaboration and empowerment.
Establishing an \textbf{open and transparent communication} channel to promote the exchange of ideas, opinions, and feedback between IT professionals and the management team creates an environment in which everyone feels heard and has the opportunity to contribute to the company's decisions. 
P13 stated: \textit{``We should listen to the talent and make him feel free to be heard, to talk, and to give ideas''}.
\textbf{Empowerment and autonomy} are aimed at delegating responsibilities to talented people so that they can make decisions within their areas of expertise. This attitude demonstrates confidence in their abilities, allowing them to feel more engaged and valued.
As P14 shared: \textit{``In my team, talents have the autonomy to make decisions. They are self-managing employees who are able to work independently and prioritize their tasks effectively''}.
Participating in the organization's \textbf{strategic projects} allows talented people to contribute to the company's success, develop new skills, and have a sense of purpose in their work. 
As P8 explained: \textit{``I encourage talented people to have challenges suited to their size and potential''}.

\subsubsection{Collaborative Work Environment} \label{subsubsec:collaborative_environment}

Talented people feel valued and engaged, have the opportunity to share and expand their knowledge and acquire a sense of belonging to a close-knit, collaborative team. This strengthens your bond with the organization and increases your motivation to stay and contribute to the company's success.
Create an \textbf{inclusive and respectful work environment} where all team members feel valued and comfortable voicing their opinions. This corporate synergy contributes to retaining talented people. 
P18 stated: \textit{``As a manager if I do not have a good working environment with respect, partnership, and team spirit, I cannot even start with other strategies. It is not easy, but this is the first step''}.

Use the various internal communication channels (i.e., e-mail, instant messaging, collaboration platforms, and virtual meetings) to \textbf{celebrate individual and collective achievements} and promote a positive and motivating climate in the work environment.
As P16 shared: \textit{``My position gives me almost no freedom to financially recognize a talent. However, I have the autonomy to publish a corporate message of congratulations so that these people feel valued''}.

\subsection{Coping Mechanisms} \label{subsec:coping}

When the factors that negatively affect talent retention are not improved, IT managers can resort to different coping mechanisms to deal with the situation. This category refers to the means or tools that can be used to deal with frustration in the talent retention process in order to establish turnaround actions.

\textbf{Job rotation} involves moving talented IT professionals to different positions or projects within the organization to allow them to acquire new skills, expand their knowledge, and have the opportunity to work in different business areas. Job rotation can help combat monotony and disinterest, keeping talent motivated and engaged. In addition, it offers opportunities for professional growth and career development, which can increase the organization's attractiveness as a place to work in the long term.
As P11 explained: \textit{``As an IT manager, you must be concerned about finding the best activities for talent. If he stopped performing well, you should plan actions to remove him from the project or area and drive him to another one''}.

\textbf{Strengthening the relationship between talent and leadership} creates a more positive, collaborative work environment where IT professionals feel valued, heard, and supported. When professionals have a healthy, trusting relationship with their leaders, they feel more motivated and committed to the organization. In addition, a good relationship between the IT team and leadership allows professionals to express their concerns, challenges, and needs, opening space for dialogue and the search for solutions together.
As P7 shared: \textit{``I think that remuneration is obviously important, but what IT managers should pay attention to is knowing how to listen to their talents, knowing what their expectations are for their career, and how we can align their goals in the organization''}.


When factors that negatively influence talent retention, such as poor compensation, cannot be readily addressed due to budget constraints or other barriers, \textbf{overtime authorization} can be a way to provide a temporary solution for IT professionals. This mechanism allows them to increase their income, even temporarily, helping to alleviate any financial pressures and providing an additional incentive to stay with the organization. However, it is important to emphasize that the authorization of overtime as a coping mechanism must be used with care. It should not be seen as a definitive solution to talent retention problems, as constantly working overtime can lead to exhaustion and burnout for professionals, negatively affecting their quality of life and well-being.
P5 stated: \textit{``I authorize overtime on a regular basis to be added to the salary. It is a way to motivate, but you have to be careful with legal rights and human fatigue''}.

\textbf{Productivity policies} (i.e., additional days off for birthdays or family events) aim to recognize and reward exceptional performance by talented people, encouraging them to continue to excel and contribute meaningfully to the organization. Hence, this mechanism contributes to the balance between work and personal life, the well-being of employees, and the retention of talent in the IT organization.
As P11 shared: \textit{``I usually authorize talented people to take a day off to extend the holiday and enjoy a longer break, as long as the activities have been delivered on time and with due quality''}.
\section{Discussion} \label{sec:discussao}


We ground our framework in two key dimensions that play a key role in retaining talent in IT organizations: satisfaction and motivation: 
\textbf{i) satisfaction:} factors identified in our framework, such as salary and benefits, professional growth and development, organizational culture, and balance between work and personal life are directly related to the satisfaction of IT professionals. When employees are satisfied with these aspects, they are more likely to stay with the organization and be committed to their work; and  \textbf{ii) motivation:} elements of our framework, such as recognition and rewards, challenging projects, and collaborative work environment are factors that can motivate IT professionals and play a crucial role in retaining talent. When organizations provide opportunities for growth, interesting challenges, and a stimulating work environment, they contribute to employee motivation and their willingness to stay with the company.

Our work complemented the studies of Gurcan and Sevik \cite{gurcan2019expertise} and Ana Maria \cite{maria2021significant} when we address talent retention specifically in the context of SE, as we provide valuable nuances, such as: \textbf{i) specific skills and expertise:} software developers possess specialized technical skills and knowledge that are in high demand across various industries; \textbf{ii) market competition:} due to the global demand for software and the rapid growth of technology companies, software developers often have more employment opportunities and job mobility than other professionals; \textbf{iii) innovation and knowledge continuity:} losing experienced software developers can result in loss of organizational knowledge and disrupt the continuity of ongoing projects; and \textbf{iv) changing technology landscape:} organizations need to retain developers who are adaptive and willing to embrace technological advancements. Software developers are indeed driven by challenges and have a strong desire for continuous learning and these roles might differ from those in other industries \cite{hall2008we}.


Moreover, we have to take into account that this work took place in a COVID-19 pandemic scenario. The pandemic brought about significant changes in the work environment and human resource management practices, which may have impacted the ways in which software engineers perceived their careers and their relationships with organizations \cite{shen2020impact}.

\subsection{Key Findings}

While a competitive salary is certainly an important factor in retaining talent in an IT organization, it is not necessarily the main factor. Talent retention is influenced by a combination of several factors, which vary according to the individual preferences and needs of IT professionals \cite{sutton2007no}.
Our results are aligned with RBV theory \cite{barney1991firm} and revealed that, despite a competitive salary, other relevant factors for retaining talent in IT organizations can be considered a source of competitive advantage as they help to attract and retain talent:

\textbf{Psychological safety.} The perception of a work environment where employees feel comfortable expressing their ideas, sharing concerns, making mistakes without fear of retaliation, and participating in open and constructive discussions, as reported in Section \ref{subsec:contextual}. A work environment in this way encourages collaboration, innovation, and continuous learning.
When people feel psychologically safe, they are encouraged to take risks and admit mistakes as endorsed by Edmondson~\cite{edmondson1999psychological}.

\textbf{Work-life balance.} Flexible schedules, remote work, or other policies that allow IT employees to reconcile their professional and personal responsibilities can make a difference in retaining them, as discussed in Section \ref{subsubsec:equilibrio_vidapessoal_profissional}.
We complemented the study of Pfeffer \cite{pfeffer2018dying}, which focuses on harmful management practices (i.e., long working hours, job insecurity, and a lack of support for well-being). 
It should be noted that during the pandemic, many organizations adopted remote work as a preventive measure. The possibility of working from home or in flexible locations may have changed IT professionals' preferences and expectations regarding the post-pandemic work model. Talent retention may be influenced by the remote work policy and the flexibility the company offers.

\textbf{Positive work environment.} A healthy and positive work environment with a strong organizational culture, collaboration between teams, and recognition and appreciation of employees can be key to retaining talent. An environment where IT professionals feel respected, motivated, and involved tends to attract and retain talent, as mentioned in Section \ref{subsubsec:collaborative_environment}.
Our findings corroborate previous research \cite{sutton2007no} showing that toxic individuals can have a negative influence on a workplace and how it affects talent retention.

\textbf{Professional growth opportunities.} IT professionals are often looking for opportunities to develop their skills, take on challenges, and advance their careers. Offering training programs, mentoring, job rotation programs, and merit-based promotions can help retain talent, as discussed in Section \ref{subsubsec:fasttrack_career}. 
According to Cascio and Boudreau \cite{cascio2010investing}, the positive effect arising from these good practices brings financial benefits to organizations, as an employer with a positive reputation that values and invests in its employees can attract customers and business partners.

\textbf{Recognition and non-financial rewards.} It includes accolades, leadership opportunities, participation in exciting projects, and public recognition for work done. The mix of these factors plays a significant role in talent retention, as explained in Section \ref{subsubsec:recompensa}. 
We had a divergence regarding a previous study \cite{twenge2010generational}. The study points out that Gen-Z and millennials are most likely to value money over non-monetary recognition as they step into entry-level jobs.



\subsection{Implications}\label{implicacoes_industria}

Our TR Framework can be useful in the daily lives of industry professionals.
\textbf{i) IT managers} could use the framework as a guide to identify the key factors that affect talent retention in the IT teams, assess the current situation and identify areas for improvement based on the framework categories, develop specific strategies to address the identified factors, context characteristics, and barriers in order to improve talent retention, and implement practices and policies based on the strategies proposed in the framework, such as recognition and rewards programs, personalized career plans, and a collaborative work environment; and
\textbf{ii) human resources analysts} could use the framework as a diagnostic tool to assess the effectiveness of existing talent retention practices, identify gaps and areas for improvement based on the framework categories and propose corrective actions, and collaborate with IT managers and other organizational leaders in implementing talent retention strategies aligned with the dimensions and strategies proposed in the framework.

To instantiate the TR Framework, 
we propose a guide grounded in the PDCA cycle (Plan-Do-Check-Act),
a widely used management model for continuous improvement~\cite{deming2018out}
The PDCA cycle can be applied iteratively over time, allowing industry professionals to constantly track and adjust talent retention actions, as follows: 
\textbf{(1) Plan:} analyze the current talent retention situation in the organization and identify the main factors or barriers;   
\textbf{(2) Do:} implement the strategies identified in the framework and  execute planned actions to improve talent retention.
\textbf{(3) Check:} collect relevant metrics on talent retention and evaluate the effectiveness of implemented strategies to identify areas for improvement; and
\textbf{(4) Act:} make adjustments to the framework and talent retention strategies as needed and implement continuous improvements to strengthen talent retention.

\section{Threats to Validity and Credibility} \label{sec:ameacas}

The reliability of the results is directly linked to the validity of the study. Every study has threats that should be addressed and considered together with the results, considering the classification proposed in \cite{runeson2012case}. 
Unlike quantitative studies, qualitative studies face a higher risk of threats to credibility than validity. The credibility and trustworthiness of qualitative research primarily depend on the diligence, thoroughness, and integrity demonstrated by the researchers throughout the process of data collection and analysis \cite{ronson1993real}. In the following, we will discuss potential threats to our work.

As \textit{external credibility}, we ensured a sufficient sample size by including 21 participants, which exceeds the recommended threshold for achieving saturation, as suggested by Guest et al. \cite{guest2006many}. Our participant group consisted of active IT managers who shared a relatively homogeneous background. It is worth noting that no new categories or concepts emerged during the last three interviews, indicating that saturation was reached. These factors provide us with confidence in the external credibility of our work.

Regarding \textit{internal credibility}, our concern is the researcher's bias. The researcher's involvement during interviews can influence participants' responses. Interactions between the researcher and IT managers may create a dynamic where participants feel pressured to provide answers that match the researcher's expectations. It is important to recognize this potential influence to ensure the objectivity of the results. To mitigate this threat, while the main coding was done by the two authors of this work, other two authors double-checked
and extensively involved with the axial coding process and the establishment of the emerging factors, barriers, strategies, and coping mechanisms. 
It is worth mentioning that three other researchers with more than 15 years of SE research and ESE supported the study design and data analysis.
In addition, we also kept an extensive audit trail in the form of complete transcripts from all participants. 


\section{Conclusion and Future Work} \label{sec:conclusao}

In this work, we presented the TR Framework, an actionable framework that addresses talent retention in IT organizations. Our work reveals that talent retention is influenced by a number of key factors, the importance of which varies according to the characteristics of the context in which IT professionals are inserted. In addition, we identified the main barriers that can negatively affect talent retention as well as effective strategies to overcome them. Finally, we discussed coping mechanisms that can be employed to deal with situations where talent retention strategies prove ineffective.



Based on our findings, we conclude that retaining talent is especially relevant for SE professionals, whose motivation is driven by technical challenges, continuous pursuit of learning, an environment that promotes innovation and collaboration, and valuing the well-being of personal life with flexible schedule policies and remote work options.
As \emph{future work}, we intend to perform some participative case studies across different IT organizations to understand how the framework is applied in practice and how results related to talent retention management vary across organizations. 


\newpage
\bibliographystyle{ACM-Reference-Format}
\bibliography{bibliography}

\end{document}